\documentstyle[12pt]{article}
\begin{document}
\title{On Continuous Ambiguities in Model-Independent Partial Wave
Analysis - I.}
\author{I.N.Nikitin* }
\date{}
\maketitle
\insert\footins{\ \ {\footnotesize *  E-mail:\ \ nikitin\_{$\,$}i@mx.ihep.su }}
\def\n{\vec n}
\def\O{\Omega}
\def\dO{\partial\Omega}
\def\ph{\varphi}
\def\half{{\textstyle{{1}\over{2}}}}
\def\rat#1#2{{\textstyle{{#1}\over{#2}}}}
\def\rea{\mbox{Re}\: }
\def\img{\mbox{Im}\: }
\def\intn{\int d^{2}n\ }
\def\la{\Bigl}
\def\ra{\Bigr}
\def\ket#1{|#1\ra>}
\def\bra#1{\la<#1|}
\def\twin#1#2{\scriptstyle #1\atop\scriptstyle #2}
\def\no{\nonumber}

\begin{abstract}
A problem of amplitude reconstruction in terms of the given angular
distribution is considered. Solution of this problem is not unique. A class of
amplitudes, correspondent to one and the same angular distribution, forms a
region in projection onto a finite set of spherical harmonics. An explicit
parametrization of a boundary of the region is obtained. A shape of the region
of ambiguities is studied in particular example. A scheme of partial-wave
analysis, which describes all solutions in the limits of the region,
is proposed.
\end{abstract}

I.\ Let a quantum system, described by a wave function $\Psi$, decay into
2 spinless particles. We are studying angular distribution of decay products
in the rest frame of initial state
$$I(\n)=|\Psi(\n)|^{2} $$
and are going to obtain the wave function $\Psi(\n)$ in the form of expansion
by
spherical harmonics $$\Psi(\n)=\sum_{lm}c_{lm}Y_{lm}(\n)\ .$$
The solution of the problem:
\begin{equation}
|\Psi(\n)|=\sqrt{I(\n)},\qquad \Psi(\n)=\sqrt{I(\n)}e^{i\ph(\n)}, \label{sol}
\end{equation}
$\ph(\n)$ is an arbitrary function of angles,
\begin{equation}
c_{lm}=\intn Y_{lm}^{*}(\n)\sqrt{I(\n)}e^{i\ph(\n)}\ . \label{cY}
\end{equation}
Substituting all possible functions $\ph(\n)$ into this formula,
one obtains the sets of coefficients $c_{lm}$, correspondent to the same
distribution $I(\n)$. In this the values of the coefficients span in the
space $\{ c_{lm}\}$ some region $\O$. Let us obtain a boundary of this
region.

We denote $(lm)=\alpha$. Let us find the variation of coefficients when
$\ph(\n)$ changes
$$\delta c_{\alpha}=\intn
 Y_{\alpha}^{*}(\n) \sqrt{I(\n)}e^{i\ph(\n)}
\ i\delta\ph(\n).$$
On the boundary all $\delta c_{\alpha}$ are orthogonal to a single vector
$N_{\alpha}$ -- normal to the boundary surface (fig.1).

We introduce a scalar product in the space of coefficients in the
following way:
\begin{equation}
(a,b)=\half\sum_{\alpha}(a_{\alpha}^{*}b_{\alpha}+a_{\alpha}b_{\alpha}^{*})=
\sum_{\alpha}\rea a_{\alpha}\rea b_{\alpha}
+\sum_{\alpha} \img a_{\alpha}\img b_{\alpha}\ .
\label{sc}
\end{equation}
$$(N,\delta c)=\rat{i}{2}
\sum_{\alpha} \intn \sqrt{I}(N_{\alpha}^{*}Y_{\alpha}^{*}e^{i\ph}
-N_{\alpha}Y_{\alpha}e^{-i\ph})\ \delta\ph(\n)=0\quad\forall\ \delta\ph(\n)$$
$$\Rightarrow\ \sum_{\alpha} N_{\alpha}^{*}Y_{\alpha}^{*}e^{i\ph}
-\sum_{\alpha} N_{\alpha}Y_{\alpha}e^{-i\ph}=0\ \Rightarrow\
e^{2i\ph(\n)}={{\sum_{\alpha}N_{\alpha}Y_{\alpha}(\n)}\over
{\sum_{\alpha}N_{\alpha}^{*}Y_{\alpha}^{*}(\n)}}$$
$$e^{i\ph(\n)}=\pm\sqrt{{{\sum_{\alpha}N_{\alpha}Y_{\alpha}(\n)}\over
{\sum_{\alpha}N_{\alpha}^{*}Y_{\alpha}^{*}(\n)}}}\qquad\mbox{or}$$
\begin{equation}
\ph(\n)=\arg\sum_{\alpha}
N_{\alpha}Y_{\alpha}(\n)\quad \mbox{($+\pi$,\ for bottom  sign).}
\label{star}
\end{equation}
This function $\ph(\n)$ is correspondent to the point, lying on the boundary of
the region  $\O$
\begin{equation}
c_{\beta}(N)=\pm\intn\sqrt{I(\n)}Y_{\beta}^{*}(\n)e^{i\arg
\sum N_{\alpha}Y_{\alpha}(\n)}.
\label{amp}
\end{equation}
The sign $\pm$ in this formula is excessive. The region $\O$ is symmetrical
in reflections about the origin, because the angular distribution does not
change in the replacement $\Psi\to -\Psi\ (c\to -c)$. For the points on the
boundary this replacement is equivalent to the
transformation $N\to -N$. In fact,
sign ambiguity in (\ref{amp}) corresponds to the replacement of outer normals
to inner ones.

Phase (\ref{star}) does not change in multiplication of all $N_{\alpha}$
by positive number. One can choose the unit normal $N_{\alpha}$:
$(N,N)=\sum\limits_{\alpha}|N_{\alpha}|^{2}=1$.
\begin{quotation}{\small
Let the unit normal $N$ be given on a surface $\Gamma$. Thereby a mapping
of the surface into unit sphere $S$ is specified:
$$\Gamma\stackrel{N}{\to}S,\quad N=N(c),\ c\in\Gamma,\ N\in S\ .$$
This mapping is called Gaussian mapping \cite{Francis}.
It possesses the property equivalent to the definition:
$$(N,dc)=0.$$
Formula (\ref{amp}) specifies a parametrization of the surface
$\Gamma=\partial\O$, the parameter is unit normal $N$. Thus $\Gamma$
is defined by the inverse Gaussian mapping
$$S\stackrel{c}{\to}\Gamma,\quad c=c(N),\ N\in S,\ c\in\Gamma\ .$$
}\end{quotation}

In order to obtain {\it a projection} of the region $\O$ in some subspace
$\{ c_{lm},\ l=0..L_{\max}\}$, one should restrict $N_{\alpha}$ in
formula (\ref{amp}) into this subspace. Values $c_{\beta}$ obtained specify
a boundary of the projection of the region (fig.2).

In order to obtain {\it a slice} of the region $\O$ by specified hyperplane
$\{ c_{lm}=c_{lm}^{(0)}\mbox{ -- fixed numbers },\ l>L_{\max}\}$,
one should solve a system of equations on~$N_{\alpha}$
$$\intn \sqrt{I}\;
Y_{\beta}^{*}e^{i\arg\sum N_{\alpha}Y_{\alpha}}=c_{\beta}^{(0)}.$$
This problem is more difficult than preceding one.

\vspace{1cm}

\underline{Example.}\ \ Let $I={{1}\over{4\pi}}$. What initial states
(except the obvious S-wave $\Psi={{1}\over{\sqrt{4\pi}}}$ ) are able to
give this distribution?

\vspace{0.5cm}

The answer:\  $\Psi={{1}\over{\sqrt{4\pi}}}e^{i\ph(\n)}$.\
The boundary of the region $\O$ is given by the formula
\begin{equation}
c_{\beta}={{1}\over{\sqrt{4\pi}}}\intn Y_{\beta}^{*}e^{i\arg\sum
N_{\alpha}Y_{\alpha}}.
\label{Iconst}
\end{equation}
Now the problem consists in clear representation of this region.

1) Projection of $\O$ in finite dimensional space $\{ c_{lm},\ l=0..L_{\max}\}$
lies inside unit sphere
$$\sum_{0}^{L_{\max}}c_{\alpha}^{*}c_{\alpha}\le
\sum_{0}^{\infty}c_{\alpha}^{*}c_{\alpha}=\intn |\Psi |^{2}=1.$$
\quad
2) Projection of $\O$ on each wave $LM$ covers a circle inside unit circle
on Argand plot:
{\small
$$\mbox{when\ }N_{lm}=\cases{e^{i\phi},&$l=L,\ m=M$\cr 0,& otherwise \cr}\quad
 \ \arg\sum N_{\alpha}Y_{\alpha} =\phi+\arg Y_{LM},$$
$$c_{LM}={{1}\over{\sqrt{4\pi}}}e^{i\phi}\intn |Y_{LM}|e^{-i\arg Y_{LM}}\cdot
e^{i\arg Y_{LM}}={{1}\over{\sqrt{4\pi}}}e^{i\phi}\intn |Y_{LM}|=$$
$$\begin{array}{r|c|c|c|c|c|c}
(L,M):&(0,0)&(1,0)&(1,1)&(2,0)&(2,1)&(2,2)\\ \hline
=\half e^{i\phi}\int\limits_{-1}^{1}d\cos\theta\ \cdot&
1&\sqrt{3}|\cos\theta |&\sqrt{\rat{3}{2}}\sin\theta &
\rat{\sqrt{5}}{2}|3\cos^{2}\theta-1|&\sqrt{\rat{15}{2}}\sin\theta
|\cos\theta |&\sqrt{\rat{15}{8}}\sin^{2}\theta\\
=e^{i\phi}\cdot&1;&\rat{\sqrt{3}}{2},&\rat{\pi}{4}\sqrt{\rat{3}{2}};&
\rat{2}{3}\sqrt{\rat{5}{3}},&\sqrt{\rat{5}{6}},&\sqrt{\rat{5}{6}}\\
&1.;&0.962,&0.866;&0.861,&0.913,&0.913\\
\end{array}$$
$$(c_{L\ -M}=c_{LM})$$
}
3) Projection on S,$\mbox{P}_{0}$-waves (a space $\{ c_{0},c_{10}\}$).

\vspace{0.3cm}

When rotating phases of all $N_{\alpha}$ by $\phi_{0}$, the phases $c_{\alpha}$
also rotate by $\phi_{0}$:
$$N_{\alpha}\to N_{\alpha}e^{i\phi_{0}}\ \Rightarrow\
c_{\alpha}\to c_{\alpha}e^{i\phi_{0}}$$
In multiplication of $N_{\alpha}$ by a positive number the $c_{\alpha}$ do not
change.

Let $N_{0}=1$. We will measure the phase of P-wave with respect to the phase
of S-wave
\begin{equation}
c_{10}\to c_{10}e^{-i\arg c_{0}}, c_{0}\to |c_{0}|.
\label{redef}
\end{equation}
Let us denote \quad $\sqrt{3}N_{10}=\rho e^{i\phi}$
$$e^{i\arg(1+\rho e^{i\phi}\cos\theta)}=
{{1+\rho\cos\theta e^{i\phi}}\over{\sqrt{1+2\rho\cos\theta\cos\phi
+\rho^{2}\cos^{2}\theta}}}.$$
{}From (\ref{Iconst})
$$c_{0}={{1}\over{4\pi}}\int_{0}^{2\pi}d\ph\cdot\int_{-\rho}^{\rho}
{{dx}\over{\rho}}\;{{1+xe^{i\phi}}\over{\sqrt{1+2x\cos\phi+x^{2}}}}\qquad
x=\rho\cos\theta$$
$$={{\sin\phi}\over{2\rho}}\int_{(-\rho+\cos\phi)/\sin\phi}
^{(\rho+\cos\phi)/\sin\phi}du\;{{1-e^{i\phi}\cos\phi+e^{i\phi}\sin\phi\cdot u}
\over{|\sin\phi |\sqrt{1+u^{2}}}}\qquad u={{x+\cos\phi}\over{\sin\phi}}$$
\begin{eqnarray}
&=&{{e^{i\phi}}\over{2\rho}}\Biggl(\sqrt{1+2\rho\cos\phi+\rho^{2}}
-\sqrt{1-2\rho\cos\phi+\rho^{2}}\label{cz}\\
&&-i|\sin\phi |\ln\;{{\rho+\cos\phi+\epsilon\sqrt{1+2\rho\cos\phi+\rho^{2}}}
\over{-\rho+\cos\phi+\epsilon\sqrt{1-2\rho\cos\phi+\rho^{2}}}}\Biggr)\quad
\epsilon=\mbox{ sgn}\;\sin\phi\nonumber .
\end{eqnarray}
Analogously
$$c_{10}={{\sqrt{3}}\over{4\pi}}\int_{0}^{2\pi}d\ph\cdot\int_{-\rho}^{\rho}
{{dx}\over{\rho^{2}}}\; x\;{{1+xe^{i\phi}}\over{\sqrt{1+2x\cos\phi+x^{2}}}}=$$
\begin{eqnarray}
&=&{{\sqrt{3}}\over{8\rho^{2}}}\Biggl(2e^{i\phi}\Bigl( (\rho+\cos\phi)
\sqrt{1+2\rho\cos\phi+\rho^{2}}\nonumber\\
&&-(-\rho+\cos\phi)\sqrt{1-2\rho\cos\phi+\rho^{2}}
\Bigr)\nonumber\\
&&-4e^{2i\phi}\Bigl(\sqrt{1+2\rho\cos\phi+\rho^{2}}-
\sqrt{1-2\rho\cos\phi+\rho^{2}}\Bigr)\label{co}\\
&&+i(3e^{2i\phi}+1)|\sin\phi |
\ln\;{{\rho+\cos\phi+\epsilon\sqrt{1+2\rho\cos\phi+\rho^{2}}}
\over{-\rho+\cos\phi+\epsilon\sqrt{1-2\rho\cos\phi+\rho^{2}}}}\Biggr)\nonumber
{}.
\end{eqnarray}
The limits
\begin{eqnarray}
\rho\to0:\ & c_{0}\to1,\ c_{10}\to0&\quad
\mbox{{\small (normal lies in the plane of S-wave)}}\nonumber\\
\rho\to\infty:\ & c_{0}\to0,\ c_{10}\to\rat{\sqrt{3}}{2}e^{i\phi}&\quad
\mbox{{\small (normal lies in the plane of P-wave)}}\nonumber
\end{eqnarray}

Fig.3a shows the projection of the region $\O$
into coordinates $(c_{0},|c_{10}|)$. Dotted band is the projection
of the boundary $\partial\O$, the projection of $\O$ is shown by sloppy
hatching. Fig.3b shows the slices of $\partial\O$ by level
surfaces $c_{0}=Const$ in projection on Argand plot
$(\rea c_{10},\img c_{10})$. When $c_{0}\to1$, the slices contract
into a point $(0,0)$ (the P-wave contribution disappears). When $c_{0}\to0$,
the slice reaches the circle with the radius $\sqrt{3}/2$. For intermediate
values $0<c_{0}<1$ the slices are of the oval shape. In this the length of
radius-vector $|c_{10}|$ changes from minimal value to maximal value
in the band, displayed on fig.3a. The slices of $\O$ by surfaces
$c_{0}=Const$ lie inside the ovals. Fig.3c shows 3-dimensional
region of ambiguity.

Expansions of the functions $\Psi(\n)=e^{i\ph(\n)}/\sqrt{4\pi}$ in common
contain infinite number of harmonics. Figures display the projections
(shades), which $\O$ casts from infinite dimensional coefficient space
in specified 2- or 3-dimensional subspaces. A contribution of the infinite
number of harmonics not involved in these figures can be evaluated.
We use the equality
$$\sum_{0}^{\infty}|c_{\alpha}|^{2}=\intn |\Psi |^{2}=1\quad\Rightarrow\quad
\sum_{\alpha\neq S,P_{0}}|c_{\alpha}|^{2}=1-{c_{0}}^{2}-|c_{10}|^{2}\ .$$
On fig.3a two circles with radii $0.95$ and $1$ are shown.
For the points of $\O$, lying between these two circles, the contribution of
``invisible'' harmonics does not exceed $10\%$ of total probability
$$0.9<{c_{0}}^{2}+|c_{10}|^{2}\leq1\quad\Rightarrow\quad
\sum_{\alpha\neq S,P_{0}}|c_{\alpha}|^{2}<0.1$$
The same region is hatched on fig.3b. On the bottom figure this region
is correspondent to a part of $\O$, lying outside a sphere with radius $0.95$.
\begin{quotation}{\small
The figures are obtained as follows. $10^{6}$ random points $(\rho,\phi)$ were
generated with uniform distribution in rectangle$$\{ 0<\rho <30,\
0<\phi <2\pi\}. $$ Coefficients $c_{0},c_{10}$ were calculated by the
expressions (\ref{cz}),(\ref{co}). Redefinition (\ref{redef}) was performed:
$c_{10}\to c_{10}c_{0}^{*}/|c_{0}|,\ c_{0}\to |c_{0}|$.
Values $(c_{0},|c_{10}|)$ were placed in 2-dimensional histogram.
Besides, 5 distributions were filled on Argand plot
$(\rea c_{10},\img c_{10})$ for $c_{0}$, lying in the intervals
$$\{ p-0.03<c_{0}<p+0.03\} \quad p=0.1,\ 0.3,\ 0.5,\ 0.7,\ 0.9$$
}\end{quotation}

4) One more symmetry of the region $\O$.

In transformations of $N_{lm}$ by matrices ${\cal D}_{mm'}^{l}$ (Wigner's
functions) the coefficients $c_{lm}$ are transformed by the same matrices
$$N_{lm}\to\sum_{m'}{\cal D}_{mm'}^{l}(\omega)N_{lm'}\ \Rightarrow\
c_{lm}\to\sum_{m'}{\cal D}_{mm'}^{l}(\omega)c_{lm'}$$
(proved by the replacement of integration parameters in (\ref{Iconst})).

Unitary matrices ${\cal D}_{mm'}^{l}$ conserve scalar product (\ref{sc}).
In ${\cal D}_{mm'}^{l}$-rotations the region $\O$ transforms to itself,
with its normals.

If one has studied in detail the region $\O$ in the considered example,
then the regions of solutions for non-isotropic distributions $I\neq Const$
can also be obtained. Formula (\ref{amp}) is linear to $\sqrt{I}$:
$$\mbox{if\quad }\sqrt{I}=\alpha_{1}\sqrt{I_{1}}+\alpha_{2}\sqrt{I_{2}},\quad
\mbox{then\quad }c_{\beta}=\alpha_{1}c_{\beta}^{1}+\alpha_{2}c_{\beta}^{2}.$$
Expanding $\sqrt{I}$ by spherical harmonics: $\sqrt{I}=\sum b_{\alpha}
Y_{\alpha}$
and using in (\ref{amp}) the multiplication theorem for spherical functions
$$Y_{\alpha}Y_{\beta}^{*}=\mbox{\ linear combination of\ } Y_{\beta}^{*}\; ,$$
we see, that the region $\O_{I\neq Const}$ can be obtained from the region
$\O_{I=Const}$ by a linear transformation with coefficients depending on the
distribution $I$.

\subsection*{ Substantial questions }

\quad 1. For the scalar product (\ref{sc}) we have chosen the Euclidean scalar
product in real space, formed by real and imaginary parts of $c_{\alpha}$.
(This scalar product is also equal to real part of the Hermitean scalar product
in complex space $c_{\alpha}$.) It is clear that the scalar product is an
additional structure, which does not influence the shape of the region $\O$.
One can easily show that the replacement of scalar product leads to some linear
transformation of the parameter $N\to \tilde N(N)$ in (\ref{amp}),
i.e. the shape of the region conserves, only its parametrization changes.
One can imagine this in the following way. The replacement of the scalar
product changes the orthogonality definition and, consequently, the definition
of the normal to the surface. When the scalar product replaces,
the field of normals $N$ on the boundary $\partial\O$ changes, but the surface
$\partial\O$ remains the same.

2. In infinite dimensional space of coefficients $c_{\alpha}$ the set $\O$
is not a region. The $\O$ is a surface, lying on a sphere
$\sum\limits_{0}^{\infty} |c_{\alpha}|^{2}=\intn I(\n)=1$.
This surface maps to the region in projections into any less space.
Expressions (\ref{cY}) and (\ref{amp}) in initial space coincide
($\O=\dO$), because any function $\ph(\n)$ (for which $e^{i\ph(\n)}$ is
quadratically integrable) can be presented in the form
$\arg\sum\limits_{0}^{\infty} N_{\alpha}Y_{\alpha}(\n)$.
Actually, we consider the projection of $\O$ in some finite dimensional space
from very beginning. In derivation of formula (\ref{amp}) the vector
$N_{\alpha}$ was restricted to this space.

3. In those points of a sphere $\n_{0}$, where
$\sum N_{\alpha}Y_{\alpha}(\n_{0})=0$, the function
$\ph_{*}(\n)=\arg\sum N_{\alpha}Y_{\alpha}(\n)$ has a break (branching point).
The correspondent wave function $\Psi_{*}=\sqrt{I}e^{i\ph_{*}}$ also breaks
in this point\footnote{However, it is single-valued, bounded and expansible by
spherical harmonics in the vicinity of $\n_{0}$. A detailed study shows that
formula (\ref{amp}) holds true for these functions.},
except a special case $I(\n_{0})=0$. This circumstance is complicated by the
fact, that all functions of form (\ref{sol}), close to $\Psi_{*}$,
also have breaks. (The phase of $\Psi_{*}$ changes by $2\pi$ in bypassing
around $\n_{0}$. All functions, close to $\Psi_{*}$, also possess this
property. If $|\Psi_{*}|^{2}=|\Psi |^{2}=I\neq0$ in the vicinity of $\n_{0}$,
then $\Psi$ should break in this vicinity.)

Therefore, $\O$ includes regions, correspondent to irregular functions.
If the regular functions $\Psi$ should be found, for which
$|\Psi |^{2}=I$ {\it exactly}, then additional analysis is needed to
discard the regions of irregularity. Outside class (\ref{sol}) smooth
functions $\Psi$ exist, which are close to $\Psi_{*}$ everywhere, except
a small vicinity of $\n_{0}$:
$$\forall \varepsilon ,\delta\ \exists\Psi\in C^{1}(S^{2})\
|\Psi(\n)-\Psi_{*}(\n)|<\varepsilon\ \forall\n :\ |\n-\n_{0}|>\delta.$$
Inside $\delta$-vicinity of $\n_{0}$ the difference $\Psi-\Psi_{*}$ is bounded.
(We present functions $\Psi_{*}=z/|z|$ and $\Psi=z/(|z|+\delta)$ as an example,
see fig.4).

The phase of $\Psi$ changes by $2\pi$ in bypassing around $\n_{0}$. From the
continuity, a point exists in $\delta$-vicinity of $\n_{0}$, where the $\Psi$
and the correspondent density $|\Psi |^{2}$ vanish. Outside $\delta$-vicinity
$|\Psi |^{2}$ can be close to $I$ as one likes. Coefficients $c_{\alpha}$
for $\Psi$ are close to the coefficients for $\Psi_{*}$. Therefore, the
distinction of $\Psi$ and $\Psi_{*}$ is not physically significant.

\vspace{0.5cm}

II.\ Let us consider a fermion, decaying into 2 particles with spins $0$ and
$\half$. The initial state is described by 2-component spinor
$$\Psi(\n)=\left(\begin{array}{c}
\Psi_{1}(\n)\\ \Psi_{2}(\n)\\ \end{array} \right) ,\qquad
\Psi_{i}(\n)=\sum_{lm}c_{lm}^{i}Y_{lm}(\n).$$
The angular distribution observed is
$$I(\n)=|\Psi_{1}(\n)|^{2}+|\Psi_{2}(\n)|^{2}.$$
\begin{quotation}{\small
The space of states is a direct product of an orbital space of decay particles
and a spin space of the final particle with the spin $\half$.

The state with definite orbital moment $l$ and projections $m,\lambda$ is:
$$\ket{lm;\half\lambda}=Y_{lm}(\n)\chi_{\lambda},\qquad
\chi_{+\half}=\left(\begin{array}{c}1\\ 0\\ \end{array}\right),\quad
\chi_{-\half}=\left(\begin{array}{c}0\\ 1\\ \end{array}\right).$$
The state with definite total moment $J$ and projection $M$ is:
$$\ket{JM}=\sum_{\twin{l=J\pm\half}{\lambda=\pm\half}}
\la< l\ M-\lambda;\half\lambda\ket{JM}\cdot \ket{ l\ M-\lambda;\half\lambda}.$$
If the coefficients of expansion of each component of the spinor $\Psi$ by
spherical harmonics are known
$$\Psi(\n)=\sum_{lm\lambda}c_{lm}^{\lambda}Y_{lm}(\n)\chi_{\lambda},$$
it is possible to obtain spin composition of the initial state
$$\tilde c_{JM}=\la< JM\ket{\Psi}=\sum_{\twin{l=J\pm\half}{\lambda=\pm\half}}
c_{l,M-\lambda}^{\lambda}\ \la< l\ M-\lambda;\half\lambda\ket{JM}^{*}.$$
}\end{quotation}
Let us find $\Psi(\n)$.
$$|\Psi_{1}(\n)|=\sqrt{I(\n)}\cos\phi(\n),\quad
|\Psi_{2}(\n)|=\sqrt{I(\n)}\sin\phi(\n),$$
$\phi(\n)$ is an arbitrary function of angles. This function defines a degree
of polarization of the final particle with the spin $\half$, which is not
measured in the experiment.
$$\Psi_{1}(\n)=\sqrt{I(\n)}\cos\phi(\n)e^{i\ph_{1}(\n)},\quad
\Psi_{2}(\n)=\sqrt{I(\n)}\sin\phi(\n)e^{i\ph_{2}(\n)},$$
\begin{equation}
c_{lm}^{i}=\intn Y_{lm}^{*}(\n)\sqrt{I(\n)}e^{i\ph_{i}(\n)}\cdot
\cases{\cos\phi(\n),& $i=1$\cr \sin\phi(\n),& $i=2$\cr} \label{ferm}
\end{equation}
One can apply to this formula the analysis described above. The result is the
following. The boundary of the region of ambiguity is parametrized by complex
numbers $N^{i}_{\alpha}$. The point on the boundary $c^{i}_{\alpha}(N)$
is defined by the formula (\ref{ferm}), where
$$\ph_{i}=\arg\sum_{\alpha}N^{i}_{\alpha}Y_{\alpha}\quad\mbox{and}\quad
\mbox{tan\ }\phi={{|\sum_{\alpha}N^{2}_{\alpha}Y_{\alpha}|}\over
{|\sum_{\alpha}N^{1}_{\alpha}Y_{\alpha}|}}.$$

\vspace{0.5cm}

III.\ Let the process be described by density matrix. We suppose that
only spinor components were mixed, i.e. only the information about
polarizations is lost. For example, consider the process:
$0+\half\to X\to 0+\half$,\quad
2 plane waves scatter in the initial state, the
initial fermion is not polarized,
the polarization of final fermion is not measured.

The initial state is
$$\ket{i\lambda_{i}}=\delta(\vec k -k_{i}\hat z)\chi_{\lambda_{i}}\sim
\sum_{l}\sqrt{{{2l+1}\over{4\pi}}}Y_{l0}(\n)\cdot\chi_{\lambda_{i}}.$$
The orbital part of the wave function $\ket{i}$ is known, the spin part is
unknown.

The initial state is described by density matrix:
$$\rho_{i}=\half\sum_{\lambda_{i}=\pm\half} \ket{i\lambda_{i}}
\bra{i\lambda_{i}}. $$
The intermediate state $X$ depends on $\lambda_{i}$:
$\ket{\Psi_{\lambda_{i}}}=M\ket{i\lambda_{i}},$ where $M$ is a reaction
amplitude.
Components of the wave function $X$ and coefficients of their expansions by
spherical harmonics represent matrix elements of reaction amplitude
$$\Psi_{\lambda_{i}\lambda_{f}}(\n)=\bra{\n\lambda_{f}}M\ket{i\lambda_{i}},\quad
\ket{\n_{0}\lambda_{f}}=\delta^{2}(\n-\n_{0})\chi_{\lambda_{f}},$$
$$c_{lm}^{\lambda_{i}\lambda_{f}}=\bra{lm;\half\lambda_{f}}
M\ket{i\lambda_{i}},\quad
\ket{lm;\half\lambda_{f}}=Y_{lm}(\n)\chi_{\lambda_{f}}.$$
The angular distribution observed is
$$I(\n)=\sum_{\lambda_{f}=\pm\half}\bra{\n\lambda_{f}}
\underbrace{M\rho_{i}M^{+}}_{\rho_{X}}\ket{\n\lambda_{f}}=$$
$$=\half\sum_{\twin{\lambda_{i}=\pm\half}{\lambda_{f}=\pm\half}}
\bra{\n\lambda_{f}}M\ket{i\lambda_{i}}\ \bra{i\lambda_{i}}M^{+}
\ket{\n\lambda_{f}}=
\half\sum_{\twin{\lambda_{i}=\pm\half}{\lambda_{f}=\pm\half}}
|\Psi_{\lambda_{i}\lambda_{f}}(\n)|^{2}$$
\begin{eqnarray}
\Psi_{\lambda_{i}\lambda_{f}}(\n)
&=&\sqrt{2I(\n)}e^{i\ph_{\lambda_{i}\lambda_{f}}
(\n)}\cdot\cases{\cos\phi_{1}(\n),& $\lambda_{i}=+\half$\cr
\sin\phi_{1}(\n),& $\lambda_{i}=-\half$\cr}\ \
\cdot\cases{\cos\phi_{2}(\n),& $\lambda_{f}=+\half$\cr
\sin\phi_{2}(\n),& $\lambda_{f}=-\half$\cr} \no\\
c^{\lambda_{i}\lambda_{f}}_{\alpha}&=&\intn Y^{*}_{\alpha}(\n)
\Psi_{\lambda_{i}\lambda_{f}}(\n)\no .
\end{eqnarray}
This formula contains 6 arbitrary functions. For the points on the boundary
these functions have a form:
\begin{eqnarray}
\ph_{\lambda_{i}\lambda_{f}}&=&\arg\sum_{\alpha}N^{\lambda_{i}\lambda_{f}}
_{\alpha}Y_{\alpha}\no\\
\mbox{tan\ }\phi_{1}&=&{{R^{-+\ 2}+R^{--\ 2}-R^{+-\ 2}-R^{++\ 2}\pm\sqrt{D}}
\over{2(R^{-+}R^{++}+R^{--}R^{+-})}}\no\\
\mbox{tan\ }\phi_{2}&=&{{-R^{-+\ 2}+R^{--\ 2}+R^{+-\ 2}-R^{++\ 2}\pm\sqrt{D}}
\over{2(R^{-+}R^{--}+R^{+-}R^{++})}}\no\\
D&=&(R^{++\ 2}+R^{+-\ 2}-R^{--\ 2}-R^{-+\ 2})^{2}
+4(R^{-+}R^{++}+R^{--}R^{+-})^{2},\no\\
\mbox{where\ }R^{\lambda_{i}\lambda_{f}}&=&
|\sum_{\alpha}N^{\lambda_{i}\lambda_{f}}_{\alpha}Y_{\alpha}|.\no
\end{eqnarray}
\underline{Notions.}

1) In view of axial symmetry of the problem the distribution $I$ does not
depend on azimuthal angle. The amplitude $\Psi$ is also proposed to be axially
symmetrical, its expansion contains harmonics $Y_{l0}$ only. Derivation of
ambiguity region in the space $\{ c_{l0}\}$ is equivalent to sectioning of $\O$
by this space. The result is described by the same expressions, in which
the indices $\alpha$ are restricted to the set $(l0)$.

2) All mentioned above corresponds to the case of inelastic
scattering (particles in initial and final states are different). For elastic
scattering it is necessary to extract explicitly in $S$-matrix a contribution
of the non-scattered wave
$$S=1+iM.$$
The consequence of unitarity of $S$-matrix is an optical theorem for
the amplitude of elastic scattering:
$$\img\Psi\bigr|_{\n=\hat z}\sim\sigma_{\mbox{\small tot}}.$$
(If the initial state is fixed, then the optical theorem is the
unique consequence
of the unitarity of $S$.) This condition define a hyperplane in the
space of coefficients. The slice of the region of ambiguity by this hyperplane
is the solution of the problem in the case of elastic scattering.

3) For pure elastic scattering the combined imposition of unitarity
and rotational invariance of $S$-matrix fixes the continuous ambiguity
(for example, see \cite{Landau} \S125). With nonelastic channels this does not
occur. Rotational invariance leads to diagonality of $S$-matrix with respect to
quantum numbers of the total moment
$$\bra{J'M'n'}S\ket{JMn}=\delta_{JJ'}\delta_{MM'}S_{nn'}^{(JM)}\ ,$$
index $n$ includes all other quantum numbers, e.g. the type of a particle.
The matrices $S_{nn'}^{(JM)}$ are unitary. This condition provides non-linear
relations among partial amplitudes of different reactions. An exact inclusion
of the unitarity condition is possible only in the
measurement and combined analysis of the whole variety of connected channels.

\vspace{1cm}

An extensive literature is devoted to the question of continuous ambiguities
in phase analysis (see review in \cite{Nikitiu},pp.291-308).
The reaction amplitude
is ambiguously reconstructed from experimentally observable quantities,
because the part of information about amplitude is lost in the measurement.
Non-observable quantities (phases, polarizations) can be functions of
angles, energy, etc., therefore the final expression for the amplitude
contains functional arbitrariness. This arbitrariness is significant
(if one substitutes fast changing phase $\ph(\n)$ in formula (\ref{amp}),
then the contribution of low harmonics decreases, the same distribution is
described by high harmonics). One can diminish the ambiguity only imposing
extra restrictions on the form of amplitude, i.e. adopting the particular
model.
Note, that in the case of the known region of ambiguity the inclusion of extra
restrictions reduces to sectioning of the region by the surface of the
restriction.
\begin{quotation}{\small
For example, it is known that the initial state $\Psi$ has definite parity.
In this case one should consider the region $\O$ in the space
$\{ c_{lm},\ l$ have the given parity $\}$. In this one can use formula
(\ref{amp}), restricting $l$ in the indices $\alpha$ and $\beta$ to the set
with the given parity. One can prove this considering the slice of $\O$ by
this space or repeating the reasonings in section I
for $\Psi$ with the definite
parity and restricting the integrations on semi-spheres.
}\end{quotation}
The state $\Psi(\n)=\sqrt{I(\n)}e^{i\ph(\n)}$ includes in general an infinite
number of harmonics. Coefficients $c_{lm}$ tend to zero at
$l\to\infty$ (fastly, if $\Psi(\n)$ is sufficiently smooth). In practical
partial wave analysis the experimental distribution is fitted by the finite
segment of the expansion: $\Psi=\sum\limits_{0}^{L_{\max}} c_{lm}Y_{lm}$,
in this the finite number of solutions has been obtained~\cite{Sadovsky}.
This happens because the segment of the expansion approximates the exact
solution, and just
for the finite number of values of parameters this approximation
is the best (fig.5).

At $L_{\max}\to\infty$ we turn back in infinite dimensional space,
where the number of solutions is infinite. Apparently, when $L_{\max}$
increases, the fit becomes unstable or new solutions appear, which
fill the region $\O$ in the limit $L_{\max}\to\infty$.

There is much speculation that the consideration of high harmonics is not
necessary, because noise exceeds a signal in high harmonics due to finite
angular resolution of the equipment and limited statistics. The contributions
of high harmonics cannot be precisely measured, but certainly this does not
mean that they are equal to zero. We have observed that weak and sharp
conditions
$$c_{lm}\to0\quad l\to\infty\qquad\mbox{and}\qquad c_{lm}=0\quad l>L_{\max}$$
have principal difference. In the first case the large ``ambiguity island''
exists\footnote{The region of ambiguity presented on fig.3 fills almost the
whole unit sphere.}. In the second case the finite number of solutions will
be found, closest to the boundary of the island. The situations are
possible when resonant structures in low harmonics can be also described by
correlated contributions of high harmonics. One can avoid this uncertainity,
if the whole region of ambiguity will be found in analysis. The general scheme
of such analysis is the following:
\begin{itemize}
\item
Energy axis of the system considered is divided in bins, the
angular distributions are filled in each bin.
These distributions are approximated by
smooth functions, e.g. spherical harmonics:
$$I(\n)=\sum_{lm}t_{lm}^{*}Y_{lm}(\n),\quad t_{lm}=\intn I(\n) Y_{lm}(\n)=
\la< Y_{lm}\ra>_{I}.$$
\item
The boundary of the ambiguity region is obtained by formula (\ref{amp}).
The integral on a sphere entering this formula can be numerically
evaluated.

{\small This integral is not expectation value of some quantity, in its
calculation the whole distribution $I(\n)$ must be known. Also, it should not
be calculated taking a root $\sqrt{I(\n)}$ from each bin content and summing
over bins. Many-dimensional distributions are rarely filled and the errors
of such integration will be great. The approximation of $I(\n)$ by the smooth
function in the previous step is necessary.}

A many-dimensional surface obtained can be kept in the form of linear frame.
A displaying technique for many-dimensional surfaces presented in this form
is being developed nowdays~\cite{NT}.
\end{itemize}
Note, that our approach does not require the fitting of the experimental
distribution. This excludes various problems, such as the derivation of
all best approximations, which requires either repeated applications of fitting
procedure with different start points or complex algebra for discovering
analytic relations between different solutions.

A presence of resonance in some wave is established independently on a model
only if the projection of ambiguity region in 3-dimensional space
$$\mbox{Argand plot for this wave}\quad\times\quad\mbox{energy axis}$$
is a narrow tube containing only resonant paths (when the energy changes,
the absolute value of the wave has a bump, the phase changes by $\pi$). In
other cases the non-resonant paths are present along with resonant ones.
In this case one may only state that the
experimental data {\it do not contradict}
the presence of the resonance in the given wave.

A model predicts the path $c(E)$ (or the class of paths). If the ambiguity
region is obtained, one can easily check, whether the path is contained in the
ambiguity region and also find the area of values of model parameters,
when the paths are contained in $\O(E)$. Particularly, these parameters
might be masses and widths of resonant states.

\vspace{0.5cm}

\hfill {\it Received December 8, 1994.}
\newpage
\subsection*{Figures captions}
\parindent=0cm

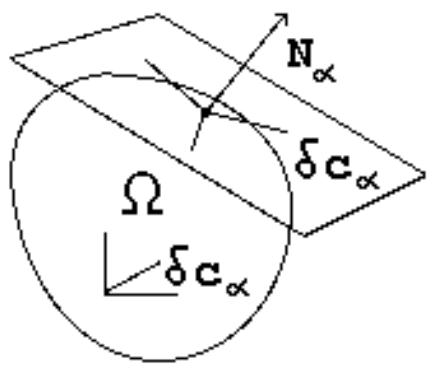
{\bf Fig.1.\ } For the point inside the region $\O$, the linear span of all
possible variations $\delta c_{\alpha}$ coincides with the whole space.
For the point on the boundary of $\O$ all $\delta c_{\alpha}$ lie in the
tangent plane to the boundary surface and are orthogonal to normal vector
$N_{\alpha}$.

\vspace{0.5cm}

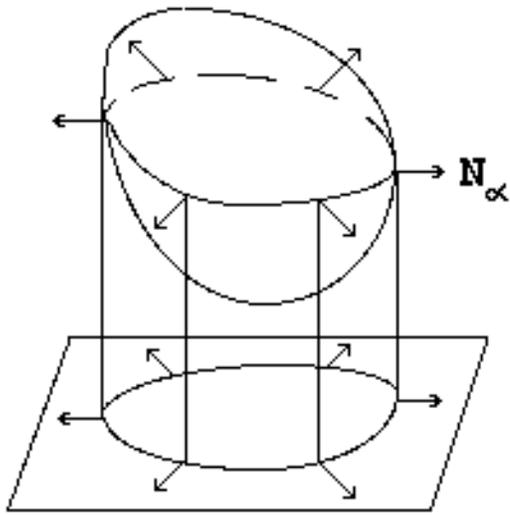
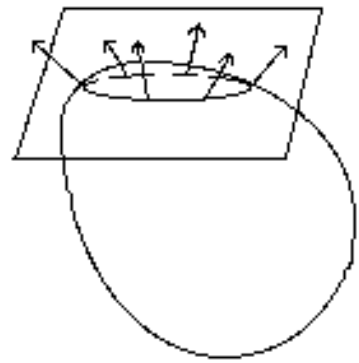
{\bf Fig.2.\ } The region $\O$, shade and slice.

\vspace{0.5cm}

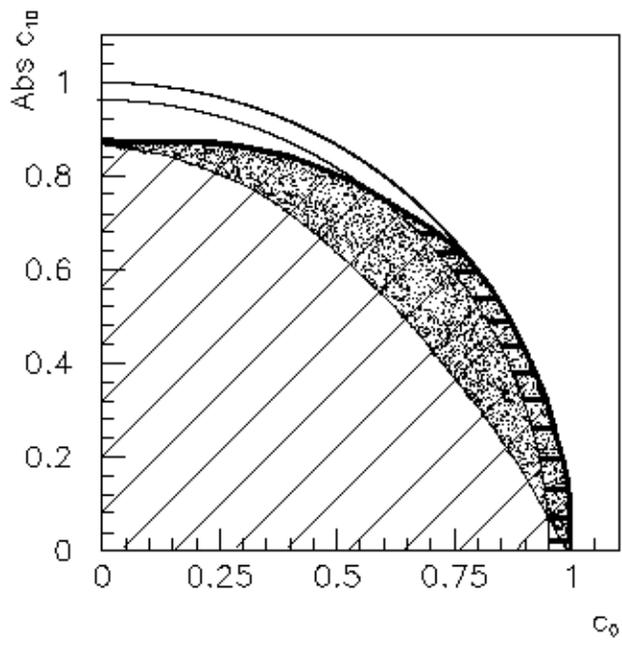
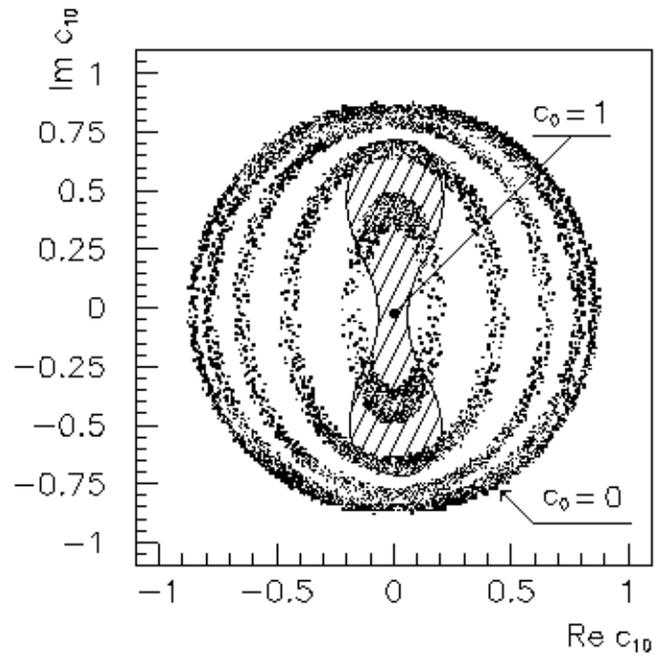
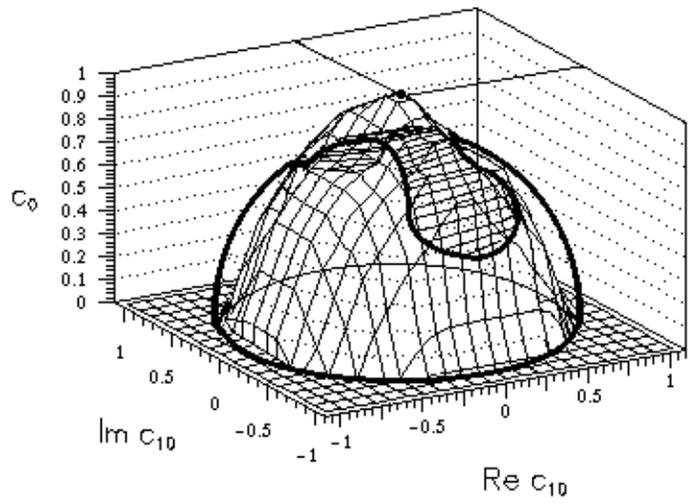
{\bf Fig.3.\ } The ambiguity region for isotropic distribution in projection
on $SP_{0}$-waves (see text for explanations).

\vspace{0.5cm}

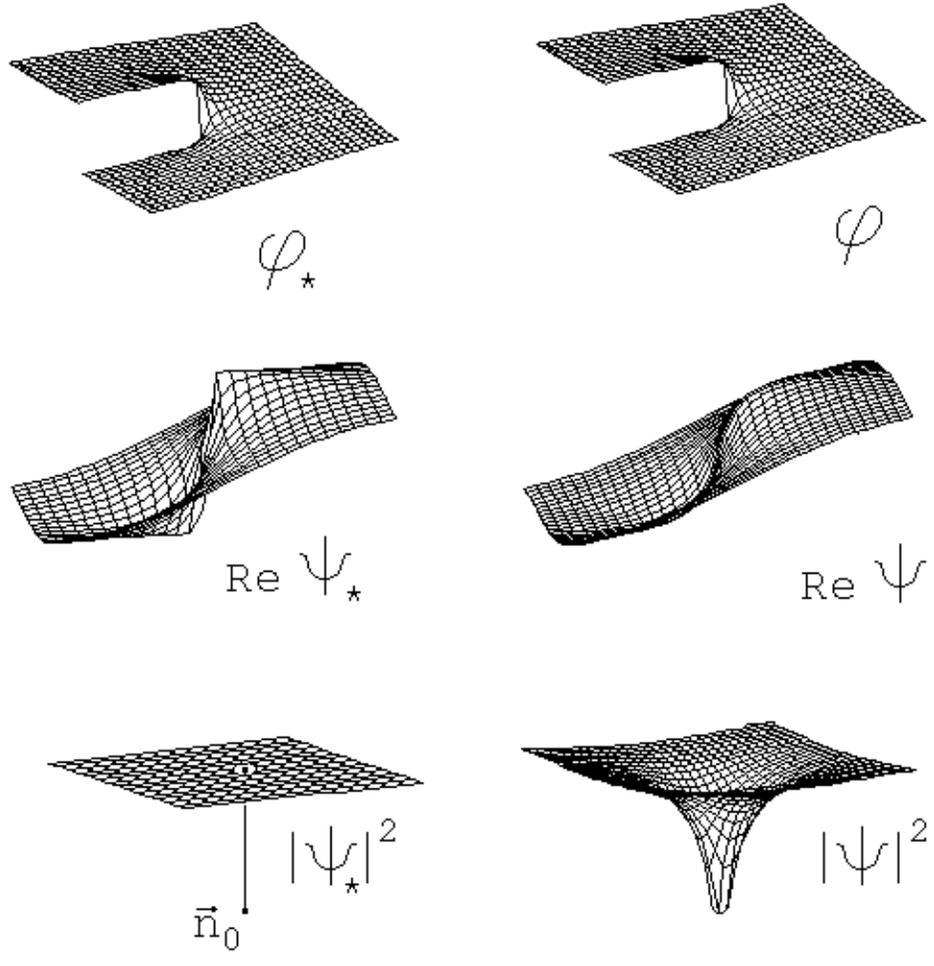
{\bf Fig.4.\ } Function $\Psi_{*}$ has a break in the point $\n_{0}$.
All functions, close to $\Psi_{*}$, for which $|\Psi |^{2}=|\Psi_{*}|^{2}$,
have breaks in the vicinity of $\n_{0}$. Continuous functions $\Psi$ exist,
which are close to $\Psi_{*}$ everywhere, except the small vicinity of
$\n_{0}$.
For these functions the $|\Psi |^{2}$ vanishes in some point inside the
vicinity.

\vspace{0.5cm}

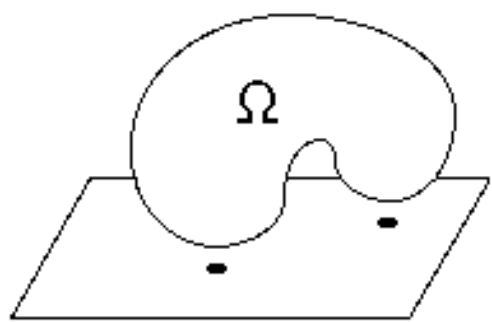
{\bf Fig.5.\ } The restriction of the
number of harmonics and model substitutions
for the amplitude lead to the fact that the solution is considered in some
narrow class of functions. This class and the set of exact solutions $\O$
in general do not intersect each other. The distance between points in these
two classes become minimal in a discrete set of points.
\end{document}